\documentclass[conference]{IEEEtran}

\usepackage{graphicx}
\graphicspath{{./Graphics/}}

\usepackage{siunitx}	% Proper typesetting of units.
\usepackage{booktabs}	% Minimally better looking tables.
\usepackage[breaklinks]{hyperref}	% Hyperlinking while allowing text links to be ``break'' across lines.
\usepackage[all]{hypcap}	% Improves hyperref's linking to floats.
\usepackage{microtype}	% Minimally looking improved typography, but reduces amount of hyphenation.

\IEEEoverridecommandlockouts

\begin{document}
\title{Non-Destructive Testing for Black Heart Cavities in Potatoes with Microwave Radiation\thanks{A version of this paper appears in \emph{\href{http://ieeexplore.ieee.org/xpl/conhome.jsp?punumber=1000454}{Proc.\ of the European Microwave Conf.}}, pp. 815--818, London, UK, October 2016. Permission to post on \href{https://arxiv.org/}{arXiv} kindly granted by the copyright holder, The \href{http://www.eumwa.org/}{European Microwave Association}.}}

%%%%%%%%%%%%%%%%%%%%%%%%%%%%%%%%%%%%%%%%%%

\author{
	\IEEEauthorblockN{Imran Mohamed\IEEEauthorrefmark{1}, 
					Richard Dudley\IEEEauthorrefmark{1}, 
					Andrew Gregory\IEEEauthorrefmark{1}, 
					Ralf Mouthaan\IEEEauthorrefmark{1}, 
					Zhengrong Tian\IEEEauthorrefmark{1}, \\
					Paul Andrews\IEEEauthorrefmark{2} and 
					Andrew Mellonie\IEEEauthorrefmark{3}
					} 
	\IEEEauthorblockA{\IEEEauthorrefmark{1}National Physical Laboratory, Hampton Road, Teddington, Middlesex, TW11 0LW, United Kingdom\\Email: imran.mohamed@npl.co.uk}
	\IEEEauthorblockA{\IEEEauthorrefmark{2}MMG Citrus Limited, The Fresh Produce Centre, Transfesa Road, Paddock Wood, Kent, TN12 6UT, United Kingdom}
	\IEEEauthorblockA{\IEEEauthorrefmark{3}Marks and Spencer, Waterside House, 35 North Wharf Road, London, W2 1NW, United Kingdom}
}

\maketitle

%%%%%%%%%%%%%%%%%%%%%%%%%%%%%%%%%%%%%%%%%
\begin{abstract}
A first investigation into the use of microwaves for the non-destructive testing 
for the presence of black heart cavities is presented. Additionally a potato's complex 
permittivity data between \SIrange{0.5}{20}{\GHz} measured using a coaxial sensor and 
the recipe for a potato phantom are also presented. Electromagnetic 
finite-difference time-domain simulations of potatoes show that changes to how 
microwaves propagate through a potato caused by a cavity can produce measurable 
changes in $\text{S}_{21}$ at the potato's surface of up to \SI{26}{\dB}. Lab-based 
readings of the change in $\text{S}_{21}$ caused by a phantom cavity submerged 
in a potato phantom liquid confirms the results of the simulation, albeit at a much 
reduced magnitude in the order of \SI{0.1}{\dB}
\end{abstract}

\section{Introduction}
Demand for high quality produce e.g.\ potatoes from retailers and their 
customers has resulted in these produce being subjected to rigorous quality 
control checks. For example, when improperly stored post-harvest without 
enough oxygen, potatoes exhibit a decay of their centres as their cells 
asphyxiate, a condition known as black heart. Current methods for 
detecting black heart in a batch of potatoes are reliant on the statistical 
sampling and manually processed destruction. 
It's estimated that these destructive tests accounts for \SI{0.5}{\percent} of the 
post-harvest wastage volume, with a net value of more than GBP~10~million.

As a means to reduce such wastage, technologies for non-destructive 
testing (NDT) and their automation have become important in recent 
decades \cite{butz_recent_2006}. Possible 
technologies capable of NDT beneath a potato's surface exist, but 
they come with either safety issues e.g.\ x-rays \cite{kotwaliwale_xray_2014}, 
or operational or throughput limitations e.g.\ ultrasound \cite{galili_ultrasonic_1993,esehaghbeygi_detection_2011}. 

An alternative is microwave imaging which thus far has yet to have its 
NDT capabilities in a food processing setting explored. In this paper we 
will present a first evaluation on the use of microwave radiation in the NDT 
for the detection of black heart cavities in potatoes. 

\section{Simulating Microwave Propagation Through Potatoes}
To begin the study, how microwaves propagate through potato with and without black hearts 
needed to be understood. This part of the study involved the measurement of a potato's 
complex permittivity, the values of which were then used in a computer simulations as well as 
in the creation of a potato ``phantom'' that would be used in further experiments in lieu of a 
real potato.

\subsection{Complex Permittivity of Potatoes}
\begin{figure}[htb]
\centering
\includegraphics[width=7.3cm]{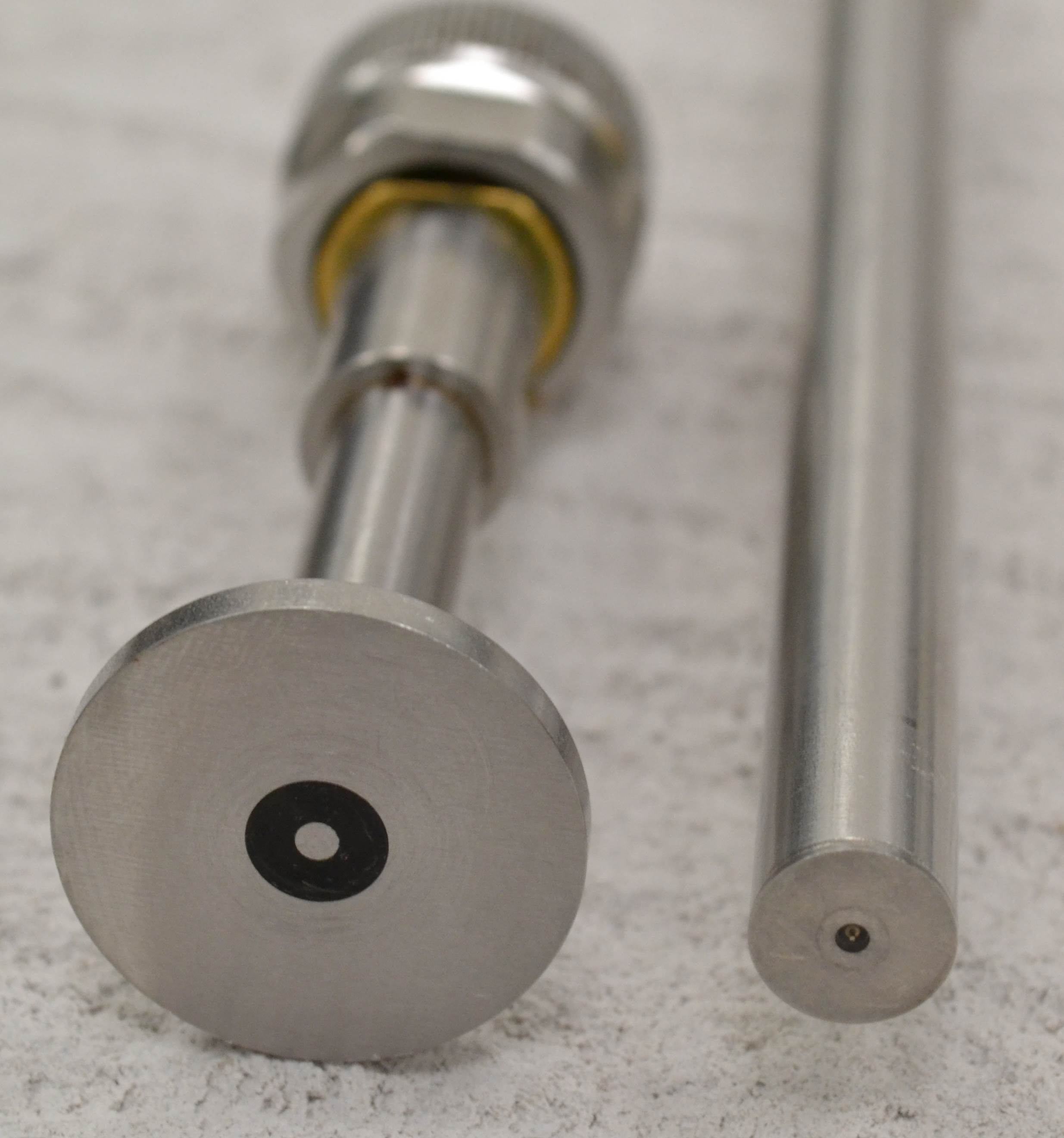}
\caption{The two coaxial sensors used during this work: NPL \SI{7}{\mm} (left) and Agilent 85070E-Perf (right).}
\label{fig:Coax-Photo}
\end{figure}

The complex permittivity of the Melody variety of potato was measured with an 
Aglient 85070-Perf coaxial sensor (Fig.~\ref{fig:Coax-Photo}) using a method detailed 
in \cite{Gregory_Dielectric_2007}. 
A potato from a store bought batch had a patch of skin removed to provide a flat 
surface for the \SI{1.6}{\mm} diameter sensor to contact. 
Calibration of the coaxial sensors were carried beforehand using the reference liquid 
procedure from \cite{Kaatze_Reference_2007}.

\begin{figure}[htb]
\centering
\includegraphics[width=7.3cm]{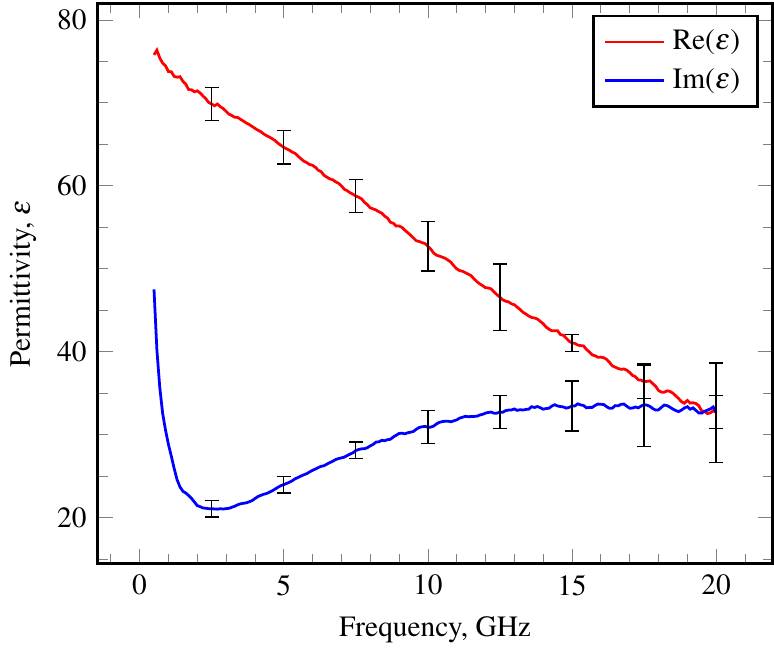}
\caption{Experimentally measured permittivity properties of Potato.}
\label{fig:potatoDE}
\end{figure}

The mean of six measurements is displayed in Fig.~\ref{fig:potatoDE}, with the 
displayed uncertainties calculated using a Monte Carlo based procedure described 
in \cite{Gregory_Dielectric_2007}. 

The data shows that both components of the complex permittivity have high 
values due to the high water content of the potato.

\subsection{Simulation Setup}
\label{sec:SimSetup}
With the measured values of a potato's complex permittivity in hand, the data was 
used in a 3D electromagnetic Finite-Difference Time-Domain (FDTD) computer 
simulation of microwaves propagating into and through a potato. The simulation 
was carried out using commercially available software (CST Microwave Studio). 
The aim of the simulation was to asses what changes the presence of a black 
heart cavity would have on the propagation of microwave radiation through a 
potato. This information would then be used to inform the design of the setup 
required for the lab-based measurements. 

\begin{figure}[htb]
\centering
\includegraphics[width=7.3cm]{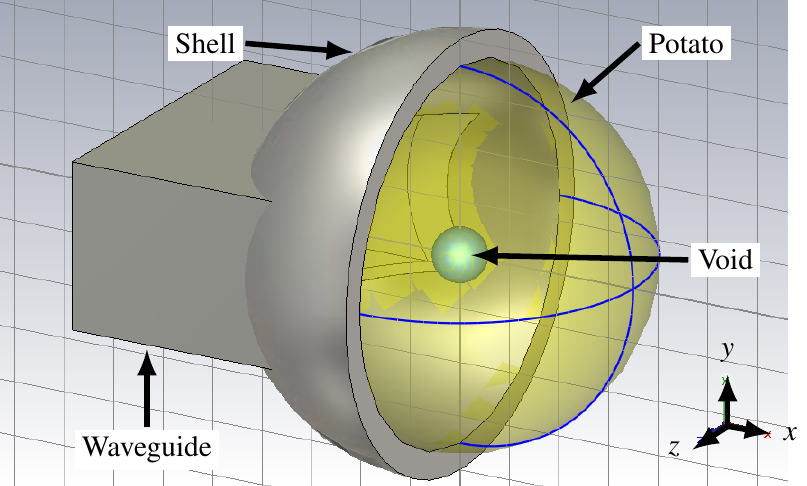}
\caption{Labelled screenshot of potato, hemispherical shell, waveguide and void from computer simulation.}
\label{fig:cstSetup}
\end{figure}

\begin{figure}[htb]
\centering
\includegraphics[width=7.3cm]{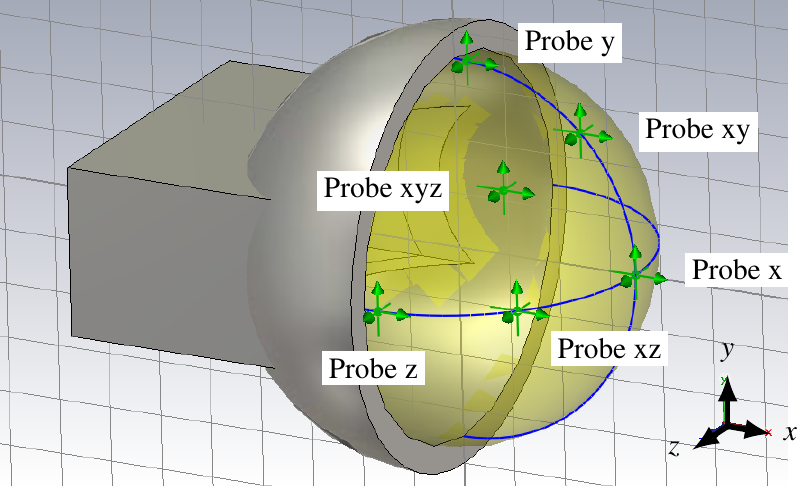}
\caption{Location and names of probes in CST simulation are marked with green triads.}
\label{fig:Probes}
\end{figure}

The simulation consists of a potato modelled as a \SI{70}{\mm} diameter sphere 
centred on the simulation space's origin (Fig.~\ref{fig:cstSetup} and 
Fig.~\ref{fig:Probes}). The black heart cavity was modelled 
as a \SI{10}{\mm} sphere with the electrical properties of a vacuum. Microwave 
radiation, linearly polarised along the $y$-axis, was projected along the 
$x$-axis into the potato with an E-band (\SIrange{3.3}{4.9}{\GHz}) waveguide. 
The waveguide's opening was drawn such that its edges matched that of 
the spherical potato's surface. In addition a hemispherical shell was added 
to the waveguide structure so that the waveguide facing side of the 
potato was encased. This was initially done to reduce the effect of surface 
waves that appeared in earlier simulations that did not have it. 
Both the waveguide and shell material's were defined as a Perfect Electric 
Conductor (PEC). 

Two sets of simulation were run. One without a black heart cavity (Fig.~\ref{fig:cstSetup}) 
and one with (Fig.~\ref{fig:Probes}). Comparison of the results from the simulations was 
carried out by looking at the electric field strength along the $xy$- and $xz$-planes of the 
simulation space and by calculating the change in measured transmission magnitude from 
six virtual probes placed on the surface of the potato's exit face i.e.\ 
$\Delta\text{mag(S}_{21}^{\text{Sim.}}) = \text{mag(S}_{21}^{\text{Void present}})-\text{mag(S}_{21}^{\text{No void}})$.

\subsection{Simulation Results \& Discussion}

\begin{figure*}[htb]
\centering
\includegraphics{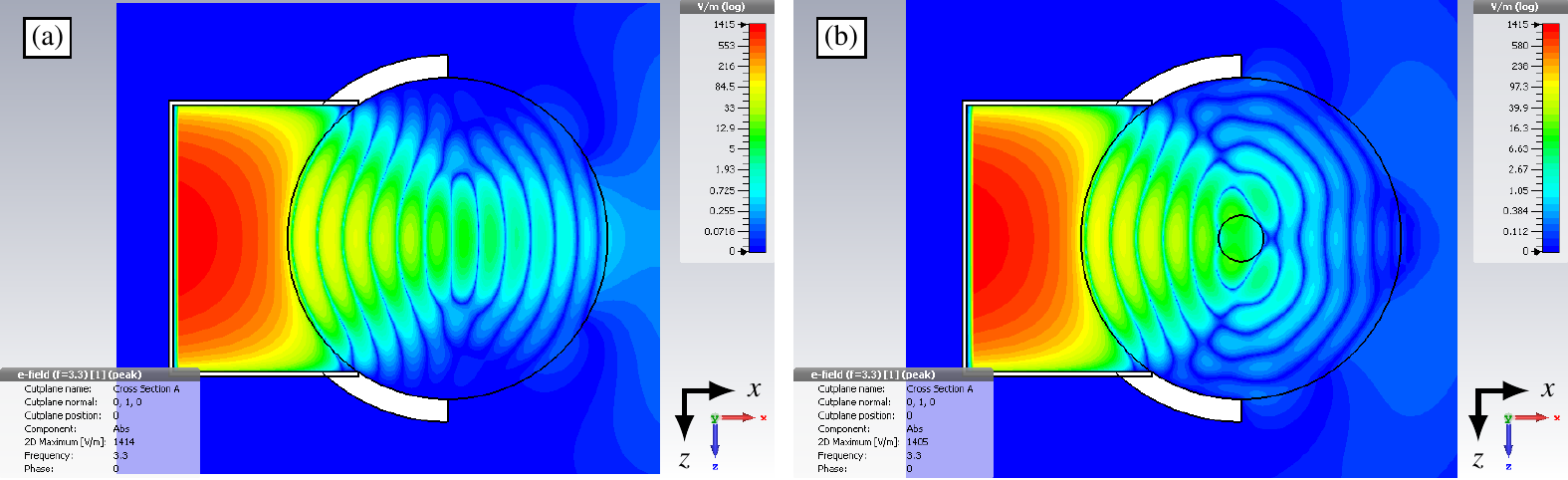}
\caption{Abs(E) at \SI{3}{\GHz} in the $xz$-plane (a) without and (b) with void.}
\label{fig:Abs-xz}
\end{figure*}

Fig.~\ref{fig:Abs-xz}
shows the simulated electric field strength 
in the $xz$- at \SI{3}{\GHz} when a black heart cavity is (a) not 
present and (b) is present. With no cavity present, the fields are seen to show that the 
potato behaves in a manner of a lossy convex lens, bringing radiation to focus whilst 
absorbing it too. 
Microwave radiation is transmitted out from the potato mainly 
from the point directly opposite the waveguide in the $xz$-plane. The presence of the 
cavity reduces the amount radiation reaching the surface at points directly behind it as 
they are cast in the cavity's shadow. Regions surrounding this point receive an increase 
in radiation due to the cavity (with its lower permittivity than the surrounding potato) 
behaving as a divergent lens, spreading radiation that wasn't reflected off its first 
interface. 

\begin{figure}[htb]
\centering
\includegraphics[width=7.45cm]{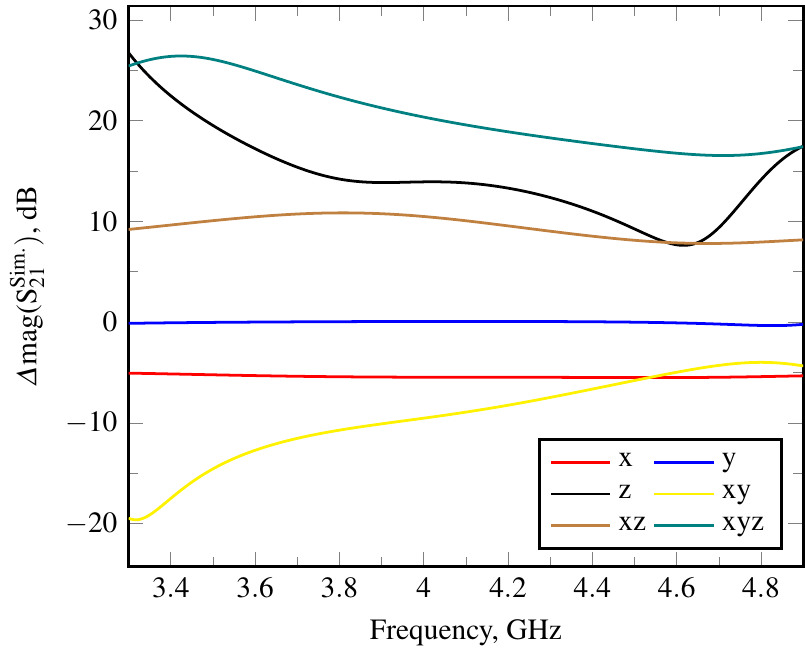}
\caption{$\Delta\text{mag(S}_{21}^{\text{Sim.}})$ measured by the probes in the computer simulations.}
\label{fig:CST-Void-1}
\end{figure}

The probe data is plotted in Fig.~\ref{fig:CST-Void-1} as 
$\Delta\text{mag(S}_{21}^{\text{Sim.}})$ against frequency across the E-band. 
As expected from the previous analysis probe~x sees a decrease of \SI{5}{\dB} across 
the band due to its location directly behind the cavity from the waveguide. 
Probes~xz, z and xyz all show increases ranging between \SIlist{10.8; 26.7}{\dB} 
across the E-band, again as expected due to the cavity spreading radiation away 
from the point directly opposite the waveguide. Probe~y sees effectively no change 
across the band, this may be due to the stronger fields in the $xy$-plane (figure not presented) 
close to the edges of the shell where probe~y is located.  

These simulation data shows that values of 
$\Delta\text{mag(S}_{21}^{\text{Sim.}})$ as large as \SI{25}{\dB} may be observed. 
The most reliably detected changes come from the probes located in the $xz$-plane. 
By comparing which probes measure an increase, decrease or no change, a cavity's 
presence and location may be deduced. Using these results, a lab-based experiment 
was designed.

\section{Measurement of $\Delta S_{21}$ in Potato Phantom}

\subsection{Potato Phantom Liquid}
\begin{figure}[htb]
\centering
\includegraphics[width=7.2cm]{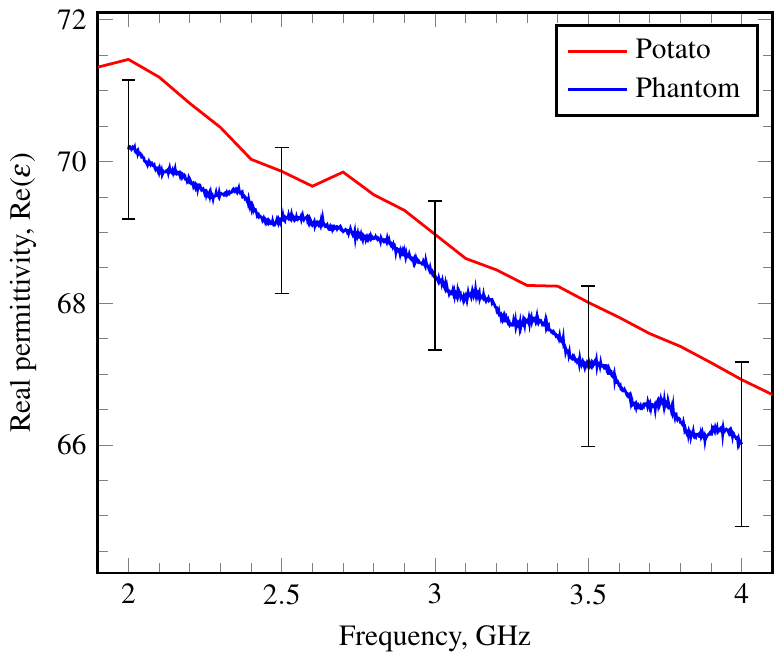}
\caption{Average of the measured real component of the potato's complex permittivity compared to that of the formulated phantom.}
\label{fig:Ralf-RE}
\end{figure}

\begin{figure}[htb]
\centering
\includegraphics[width=7.3cm]{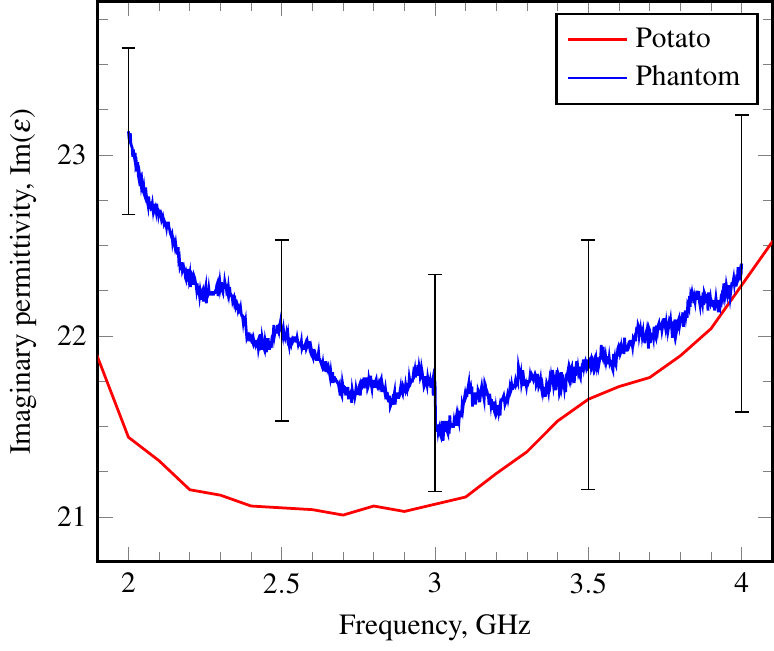}
\caption{Average of the measured imaginary component of the potato's complex permittivity compared to that of the formulated phantom.}
\label{fig:Ralf-IM}
\end{figure}

The need for a potato phantom was to allow the repeatable measurement of an object 
of known shape and size, that also exhibited the dielectric properties of a potato at 
the observed frequencies. Real potatoes suffer from dehydration over time changing 
their internal properties, in addition to coming in a variety of sizes and irregular shapes.
A solution consisting of, by weight, \SI{6.8}{\percent} polysorbate~80, \SI{1.1}{\percent} sodium chloride and 
\SI{92.1}{\percent} deionised water was created to perform as a potato phantom liquid. 
Measurements with the \SI{7}{\mm} NPL coaxial sensor (Fig.~\ref{fig:Coax-Photo}), using 
the procedure of \cite{Gregory_Dielectric_2007} show that it was successful in 
reproducing the complex permittivity of the potato between \SIrange{2}{4}{\GHz} 
(Fig.~\ref{fig:Ralf-RE} and Fig.~\ref{fig:Ralf-IM}). The agreement between the 
real and imaginary components of the potato phantom and the measured potato 
values is within \SIlist{1.2; 10}{\percent} respectively.

\subsection{Experimental Design \& Setup}
\begin{figure}[htb]
\centering
\includegraphics[width=7.3cm]{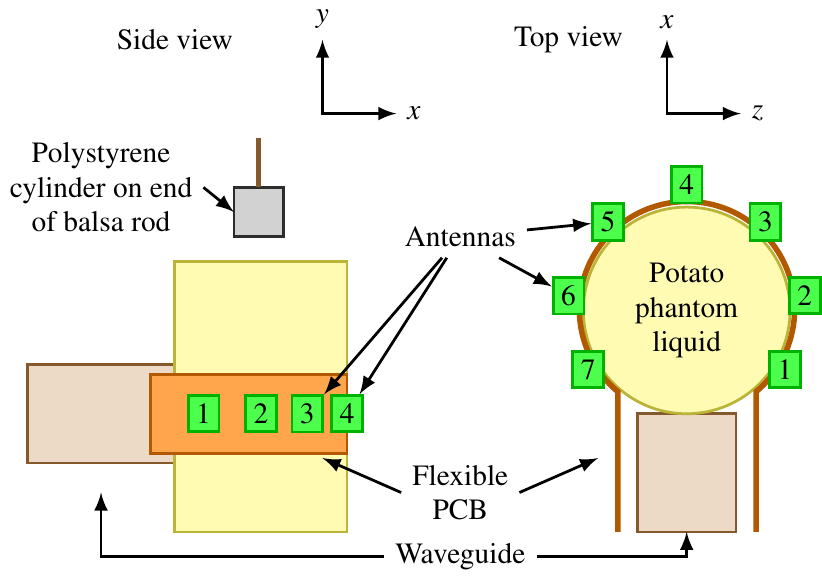}
\caption{Diagram of experimental setup. Not shown are the PIN switch which is connected to the flexible PCB, and the VNA which is connected to the waveguide and the PIN switch.}
\label{fig:labSetup}
\end{figure}

To attain repeatable measurements, the setup shown in 
Fig.~\ref{fig:labSetup} was designed, based on the setup used in 
computer simulation of Section~\ref{sec:SimSetup}. In this a 
waveguide transmits microwaves into a HDPE plastic bottle 
(external diameter: \SI{60}{\mm}, internal diameter: \SI{58}{\mm}) that is 
filled with potato phantom liquid from the previous subsection. At 
the same height as the waveguide, wrapped around the bottle's 
remaining circumference, a flexible printed circuit board (PCB) with 
seven equidistantly separated \SI{3.3}{\GHz} Murata SMT antennas 
mounted on are used to measure the microwaves that were able 
to transmit through the bottle. An expanded polystyrene cylinder 
\SI{22}{\mm} in height and \SI{22.4}{\mm} in diameter was used as a black 
heart cavity phantom.

Several design choices and changes from the simulation setup were 
incorporated based on the results of the previous section. 
Firstly, the waveguide was designed to be dielectrically loaded with 
Macor\textsuperscript{\textregistered} ceramic glass. This was to reduce 
the waveguide's dimensions to \SI{10.1 x 23.2}{\mm} and to 
improve impedance matching between it and the potato. 
Secondly, it was decided to only look for changes in the microwave 
transmission in the same plane as the waveguide. Thirdly, for practical 
reasons the potato is represented as a cylinder rather than a sphere.

To obtain a reading, a transmission reading is taken at the antennas without 
the cylinder submerged in the phantom liquid, 
$\text{mag(S}_{21}^{\text{No cyl.}})$, and another with it submerged in 
the bottle's centre to the depth of the waveguide and antennas, 
$\text{mag(S}_{21}^{\text{Cyl. present}})$. A single reading 
of is then $\Delta\text{mag(S}_{21}^{\text{Exp.}})=\text{mag(S}_{21}^{\text{Cyl. present}})-\text{mag(S}_{21}^{\text{No cyl.}})$. Measurements were 
taken with an Anritsu MS2028C vector network analyser (VNA). 

\subsection{Experimental Results \& Discussion}
The mean readings of $\Delta\text{mag(S}_{21}^{\text{Exp.}})$ from the 
seven antennas and the standard deviations are given in Table~\ref{tab:11th-Sept}.
As expected the antennas located most behind the cylinder, i.e.\ \numrange{3}{5}, 
observed a decrease in received transmission when cylinder was present. However 
this was at a level much lower than expected based on the simulation 
results, \SI{-0.17}{\dB} for antenna 4 compared to \SI{-5}{\dB} for probe~x. The remaining 
antennas were expected to see a positive $\Delta\text{mag(S}_{21}^{\text{Exp.}})$ 
due to reflected microwaves. Indeed that is seen in antennas \numlist{6;7}, measuring an 
increase of \SI{0.13}{\dB} and \SI{0.10}{\dB} respectively. Antennas \numlist{1;2} see a decrease 
however. 

\begin{table}[htb]
\centering
\caption{Measured $\Delta\text{mag(S}_{21}^{\text{Exp.}})$ by Antennas \numrange{1}{7} at \SI{3.3}{\GHz}}
\label{tab:11th-Sept}       
\begin{tabular}{SSS}
%\hline\noalign{\smallskip}
\toprule
{Antenna \#} 	& 	{Mean $\Delta\text{mag(S}_{21}^{\text{Exp.}})$, \si{dB}}	& {S.D.\ $\Delta\text{mag(S}_{21}^{\text{Exp.}})$, \si{dB}}\\
\midrule
1	&	-0.15 	& 	0.01	\\
2	&	-0.23 	& 	0.01	\\
3	&	-0.03		&	0.04	\\
4	&	-0.17		&	0.01	\\
5	&	-0.08		&	0.03	\\
6	&	0.13		&	0.02	\\
7	&	0.10 		& 	0.06	\\
\bottomrule
\end{tabular}
\end{table}

These first results demonstrate that detection of a phantom black heart 
cavity is possible by looking at the change in detected transmission levels 
of microwaves. 
The smaller than expected $\Delta\text{mag(S}_{21}^{\text{Exp.}})$ 
could be placed down to losses in the lab-based setup. Mismatch  
between the curvatures of the waveguide and bottle limits how well 
the microwaves are transmitted into the bottle. The SMT antennas are 
also not \SI{100}{\percent} efficient in receiving microwave radiation in the same 
way as the virtual probes in the simulation are. Further reductions may 
be accounted for from losses in the cables and PIN switch.
This asymmetry in the readings of antennas, \numlist{1; 2}, and \numlist{6; 7}, may 
be caused by misalignments in the setup.  

Future work will look into the measurement of the microwave's phase 
component to detect presence of black heart cavities as well as making 
improvements to the setup. 

\section{Conclusions}
In this paper we describe the first attempt of the use of microwaves for 
the non-destructive testing of black heart cavities in potatoes. Initially the 
complex permittivity of the Melody variety of potato was measured, from 
which the values were used in an EM FDTD simulation of a spherical 
potato without and with a cavity. The simulation data showed that changes 
in the amount of radiation reaching the surface between the two should allow 
the detection of a cavity. Lab-based measurements using a potato and cavity 
phantom also saw changes, albeit at a lower level than the simulations.

%%%%%%%%%%%%%%%%%%%%%%%%%%%%%%%%%%%%%%%%%%
\section*{Acknowledgements}
This work was funded by Innovate UK and National Measurement System, BIS.
Flexible PCB was manufactured by Ahmed Razak at Quantek Ltd.

%%%%%%%%%%%%%%%%%%%%%%%%%%%%%%%%%%%%%%%%%%

\end{document}